\documentstyle[12pt]{article}

\large
\newcommand{\be}{\begin{eqnarray}}
\newcommand{\ee}{\end{eqnarray}}

\begin{document} 
\setlength{\baselineskip}{23pt}
\setlength{\baselineskip}{27pt}
\pagestyle{empty}  
\renewcommand{\thefootnote}{\fnsymbol{footnote}}
\centerline{\bf\LARGE The Chiral Phase Transition}
\centerline{\bf\LARGE and Instanton$-$Anti-instanton Molecules}
\vskip 1cm
\centerline{\bf T.~Sch\"afer\footnote{supported in part by the 
  Alexander von Humboldt foundation.},
  E.V.~Shuryak and J.J.M.~Verbaarschot}
\vskip 1cm
\centerline{\it Department of Physics}
\centerline{\it State University of New York at Stony Brook}
\centerline{\it Stony Brook, New York 11794, USA}
\vskip 1cm
     
\setlength{\baselineskip}{16pt}
\centerline{\bf Abstract} 
In this paper we explore the idea that the chiral phase transition
in QCD can be described as a transition from a disordered instanton
liquid to a strongly correlated phase of polarized instanton
anti-instanton molecules. We calculate the degree of polarization
of the molecules as a function of the temperature and show that
the resulting $T$-dependence of the fermion determinant drives the
chiral phase transition. 
We also show how the polarization of the molecules can lead to
a non trivial behavior of the energy density and pressure.
 Finally, we study the effect of the presence of molecules
on the propagation of quarks at  at $T\sim
T_c$. We derive the
corresponding effective interaction and find that the strength in 
the scalar-pseudoscalar channel is four times the strength in 
the vector-axial-vector channel which agrees with recent lattice
QCD simulations. We give results for the
quark condensates as well
as mesonic and baryonic correlation functions and find that the
'screening masses' of chiral partners become equal
for $T>T_c$, where we still observe substantial attraction
in the scalar-pseudoscalar meson channels.
\vfill
\begin{flushleft}
SUNY-NTG-94-24\\
May 1994
\end{flushleft}
\eject
\newpage
\setlength{\baselineskip}{23pt}
\pagestyle{plain}
\renewcommand{\thefootnote}{\arabic{footnote}}
\setcounter{footnote}{0}
\setcounter{page}{1}

\section{Introduction} 

   Since the first suggestion (see \cite{chibreaking} and references 
therein) that instantons are related to the breakdown of chiral 
symmetry in the QCD vacuum, significant efforts were made 
\cite{Shuryak_82,DP_chi,Shuryak_88,Shuryak_Verbaarschot} 
to transform this idea into a quantitative argument.

Recently, two results have led to significant progress in connection 
with these efforts. First, it was shown \cite{RILM,wave} that even the 
simplest possible instanton-based model of the QCD vacuum, the so called
Random Instanton Liquid Model (RILM), predicts correlation
functions which agree both with phenomenological information (see the 
recent review \cite{Shuryak_cor}) and lattice calculations \cite{Negele}.
In particular, it was shown that many hadrons (including, e.g. pions 
and  nucleons) are bound by the instanton-induced interactions, and that
their properties (masses, coupling constants and wave functions) are 
reproduced by the model. 

Secondly, important new results were obtained from the study of
'cooled' lattice configurations. 'Cooling' is a procedure that
relaxes any given gauge field configuration to the closest 'classical 
component' of the QCD vacuum. As emphasized in earlier works, the 
resulting configurations were found to be of multi-instanton type
\cite{cooling}. The recent work by Chu et al. \cite{Chu_etal_2}
now concludes that the typical instanton separation is given by $R\simeq
1.1$ fm while  the typical size is about $\rho\simeq 0.35$ fm. 
These numbers essentially reproduce the key parameters  of the 
'instanton liquid' picture of the QCD vacuum, originally suggested
on purely phenomenological grounds \cite{Shuryak_82}.

 In addition to that, it was found that correlation functions 
as well as hadronic wave functions in most channels remain 
practically unchanged after 'cooling'. This confirms that the 
agreement of previous lattice calculations with the instanton model 
was not accidental. In fact, these works provide a decisive experiment
as far as the validity of the instanton model is concerned, 
demonstrating that even after eliminating such QCD phenomena 
as the  perturbative gluon exchange and confinement, one still observes
hadronic bound states. Moreover, the corresponding  masses and wave 
functions appear to be only mildly affected!

Although many details remain to be worked out, it seems 
fair to say that instantons are indeed a major component of the QCD 
vacuum, producing the quark condensate, the low lying hadronic 
states and many other nonperturbative features of QCD.

  In this paper we discuss properties of the instanton ensemble at finite 
temperature. Our focus is especially on the region around  the chiral
phase transition $T\simeq T_c$. The first attempt to understand this 
phase transition as a rearrangement of the instanton liquid, going from 
a random phase at low temperatures to a strongly correlated 'molecular' 
phase at high temperatures was made in \cite{IS}. The essential point
is that while individual instantons are strongly suppressed in the 
chirally restored phase (because of the corresponding zero modes),
instanton$-$anti-instanton ($I\bar I$) pairs have a finite probability 
even above $T_c$. This idea was recently reexamined in \cite{IS2}, and although 
the basic ingredients of the model are the same, the two works suggest 
significantly different scenarios for the chiral phase transition.

  According to the original idea, chiral symmetry is restored because
most of the instantons are removed by a finite temperature instanton 
suppression factor \cite{Shuryak_conf,PY}. This factor describes the
effect of Debye screening on the fluctuations around the instanton 
solution. Of course, it also affects the high temperature molecular phase 
and allows only a small number of molecules above $T_c$. 
In the more recent work \cite{IS2}, 
on the contrary, it was argued that screening
should not be a dominant effect at temperatures below or just above
$T_c$ (see also \cite{dens}). Instead, it was shown that the temperature 
dependence of the fermion determinant
can provide chiral restoration even without any additional suppression 
factors\footnote{In earlier works on the subject 
\cite{IS,Diakonov_Mirlin,NVZ} the high temperature 
instanton suppression factor was extrapolated to all temperatures.}. 
This means that one may have a significant number of instantons 
even above $T_c$, and that the cause of the phase transition is not the
suppression of instantons but a rearrangement of the 
instanton liquid.

  This new scenario for the chiral phase transition implies a 
number of nonperturbative effects in the region just above 
$T_c$. One of these effects was already mentioned in 
\cite{IS2}: instantons can provide a substantial contribution
to the energy density and pressure even above the chiral phase
transition. In this paper, we want to study the effects of the 
rearrangement of the instanton vacuum in more
detail. In particular, we consider the behavior of condensates and 
hadronic correlation functions through the chiral phase transition.
For this purpose we consider a schematic model in which the
instanton liquid is described as a mixture of a random and
a molecular component. The phase transition is then studied
as a function of the fraction of correlated instantons. 

 The paper is organized as follows. In section 2 we give a general
discussion on correlations in the instanton liquid and introduce
the concept of instanton$-$anti-instanton molecules. In section 3
we study the phenomenon of polarization of molecules at finite 
temperature and in section 4 we describe how this effect 
determines the energy density and pressure of the instanton
ensemble. A general treatment of the new interaction induced by
the presence of molecules is given in section 6 while in section
7 we present the results of a simulation of a mixed ensemble of
random instantons and molecules.

\section{Correlations in the instanton ensemble}

  Let us introduce the necessary general formulae and notations.
The ensemble of interacting instantons is described by a
partition function of the type
\be
\label{Z}
Z= \sum_{N_+ N_-} {1 \over N_+ ! N_- !}\int
   \prod_i^{N_+ + N_-} [d\Omega_i\; d(\rho_i) \rho_i^{N_f} ]
   \exp(-S_{int})\prod_f^{N_f} \det(\hat D+m_f) \ \ \,
\ee
where $N_+$ and $N_-$ are the numbers of instantons and anti-instantons,
$d\Omega_i$ is the measure in the space of collective coordinates
(12 per instanton in $SU(3)$) and 
\be
\label{idens}
d(\rho) &=& C_{N_C} \rho^{-5}\left( \frac{8\pi^2}{g(\rho)^2} \right)^{2N_c}
     \exp\left(-\frac{8\pi^2}{g(\rho)^2}\right) \\
C_{N_c} &=& \frac{ 4.6\exp(-1.86 N_c) }
                 {\pi^2 (N_c-1)! (N_c-2)! } 
\ee 
is the semiclassical amplitude for a single instanton which depends
on the running coupling constant $g(\rho)$. Furthermore, 
$S_{int}$ is the gluon-induced interaction, which we do not specify 
in this work. Instead, we focus on the last factor, which appears 
after the integration over fermions has been carried out. 
As is the case for the bosonic zero modes, the integral over the fermion
zero modes has to be performed exactly.
Assuming that the fermion determinant factorizes in a zero mode and
a nonzero mode factor, the first factor is given by
$\det(T T^\dagger + m_f^2)$, whereas the second factor is taken into account
to gaussian order. The matrix 
$T_{I\bar I}$ is the $N/2\times N/2$ 'hopping' matrix 
\be
\label{overl}
 T_{I\bar I} &=& \int d^4x\;  \phi_{\bar I 0}^\dagger (x-z_{\bar I})  
 i\hat D_x  \phi_{I0}(x-z_I). 
\ee
Here, $z_I,z_{\bar I}$ are the centers of the  instantons and for 
simplicity we assume that\footnote{Recall that any condition applied to a 
macroscopically large system should not be important anyway.}, 
$N_+=N_-=N/2$. The statistical mechanics described 
by the partition function (\ref{Z}) is complicated and 
so far direct simulations have only been carried out at zero
temperature.

   The instanton solutions and the corresponding zero modes are known 
analytically for non-zero temperature \cite{HS_T}, and detailed 
studies of the temperature dependence of the 'hopping' matrix elements 
were performed in \cite{SV_T,khoze}. The general structure of these matrix 
elements is given by $T_{I\bar I}=u_4 f_1+ (\vec u\vec  r/ r) f_2$.
Here, the fourvector $u_\mu=(\vec u, u_4)$ is defined by $U_{I\bar I}=
u_\mu\tau_\mu^+$, where $U_{I\bar I}$ is the upper $2\times 2$ corner
of the $N_c\times N_c$ matrix that describes the 
relative orientation of the instanton and the anti-instanton
and $\tau_\mu^+ = (\vec \tau,-i)$. The 
two invariant functions $f_1$ and $f_2$ can be parameterized in the 
form \cite{SV_T}
\be
f_1 &=& i\frac{ \pi T \sin\left( \pi\tau T \right)
               \cosh\left(  \pi r   T \right) }
             { ( \cosh\left(\pi r   T \right) 
               - \cos\left( \pi\tau T \right) + \kappa_1^2(T) )^2}
             \cdot F_1(\tau,r,T),   \\
f_2 &=& i\frac{ \pi T \cos\left( \pi\tau T \right)
               \sinh\left(  \pi r T \right) }
             { ( \cosh\left(\pi r T \right) 
               - \cos\left( \pi\tau T\right) + \kappa_2^2 (T) )^2}
             \cdot F_2(\tau,r,T). 
\ee
where $r$ and $\tau$ are the separation between the instanton and the 
anti-instanton in the spatial and time direction. All quantities 
are given in units of the geometric mean of the two instanton
radii  $\sqrt{\rho_I\rho_A}$. The functions $F_1,\,
F_2$ and $\kappa_1,\,\kappa_2$ provide the correct normalization in 
the limits of zero and infinite temperature. They are explicitly 
specified in \cite{SV_T} and used in our simulations but their 
detailed structure is not important for the purpose of our arguments.

  Neglecting all interactions one arrives at a random ensemble of
instantons in which the  distribution over all collective coordinates 
such as positions and orientation is given by the corresponding invariant
measure. As was mentioned 
in the introduction, this model leads to a simple and phenomenologically
successful description of the QCD vacuum. Notable exceptions are those
channels in which instantons produce a strong repulsion, in particular
the $\eta'$ and $\delta$ (scalar-isovector) channels. 

   Correlations in the instanton liquid have two aspects, which 
we will refer to as long and short range correlations.
Long range correlations are related to the phenomenon of 
screening of the topological charge. Its most important
manifestation is the vanishing of the global topological charge 
(and the vanishing of the topological susceptibility), 
which takes place provided the theory has massless fermions
and one is considering a sufficiently large volume. As expressed 
by the Witten-Veneziano relation, these correlations are related to the  
physics of $\eta'$ meson. We plan to discuss this question in a 
separate publication \cite{SV_screening}.

  In the present work we want to focus on short range correlations, 
which can be studied by considering $I\bar I$ pairs. 
Correlations in the instanton liquid that may lead to the formation 
of $I\bar I$ pairs are caused by  the  gluonic action $S_{int}$ 
or the effective fermion action $S_f=N_f {\rm Tr} \log (\hat D)$.
Naturally, their relative role depends strongly on the number of light 
flavors $N_f$ in the theory. However, both types of interaction
favor the same relative orientation of the instanton and the 
anti-instanton. This orientation is most simply described in terms 
of an angle $\theta$  defined by
\be 
\cos(\theta) = \frac{|u\cdot R|}{|u|\sqrt{R^2}}
\ee
where $R_\mu=(\vec r,\tau)$ is the vector connecting the centers of 
the instantons and $u_\mu$ is the relative orientation vector introduced
above. Maximal attraction, for both the gluon and the fermion 
interaction, corresponds to $\theta=0$. 

Since the probability of a given configuration is proportional to the value
of the fermion determinant, configurations with a large  determinant 
should be  preferred. For a single $I\bar I$ pair  we have 
$\det(\hat D)\sim (\cos\theta)^{2N_f}$. If this $\bar I I$  pair 
sits in the vicinity of $\theta=0$ we will refer to it as a 'molecule'. 
Clearly, it maximizes the fermion determinant,
or minimizes the corresponding 'energy', $-\log(\det \hat D)$. 

  However, dealing with a statistical system one should not simply
minimize the energy, but rather take into account the competition 
between minimizing the energy and maximizing the entropy (i.e.~minimize 
the free energy). In other words, we have to determine the relative weight 
of the most attractive (molecular) orientation relative to all others
in the $SU(3)$ group from summing over all states in the partition sum.

  Before we consider this problem, let us  make a brief digression.
Even if the contributions of 'molecules' in the QCD vacuum is not large,
there are external parameters that  can increase their role.
One way of doing this is to consider QCD-like theories with an  
increasing number of light flavors $N_f$. As the fermion determinant 
is raised to a higher power $N_f$, the role of correlations induced
by the determinant certainly increases. Thus, one may anticipate the 
existence of some critical number of flavors $N_{f,\,cr}$ above which 
the instanton ensemble is dominated by molecules 
\cite{sh-mol,Shuryak_Verbaarschot}.
In this paper, however, we want to consider another external 
parameter that has a significant effect on 
correlations in the instanton liquid, namely the temperature.

\section{Polarization of the $I\bar I$ molecules }

   In reference \cite{IS2} a schematic model was developed in order
to deal with the complicated statistical mechanics described by
the partition function (\ref{Z}). In this model the instanton
ensemble is described as a mixture of a molecular and a random
component. The differential activities for the two components
are assumed to be
\be
\label{molec}
 dZ_m&=& C^2d\rho_1d\rho_2d^4RdU\,
      (\rho_1\rho_2)^{b-5}\exp\left[ -\kappa (\rho_1^2+\rho_2^2)
      (\overline{\rho_a^2}n_a + 2\overline{\rho_m^2}n_m)\right]
      \, \langle\left( T_{I\bar I}T^*_{\bar II}\right)^{N_f}\rangle 
\ee
for the molecular component and
\be
\label{random}
 dZ_a&=& 2C d\rho d^4RdU\, \rho^{b-5}
       \exp\left[ -\kappa \rho^2
      (\overline{\rho_a^2}n_a + 2\overline{\rho_m^2}n_m)\right]
      \, \langle \frac 1N{\rm Tr}TT^\dagger \rangle^{N_f/2}
\ee
for the atomic component. Here, $n_a,n_m$ denote the densities 
of the random and the molecular components, $\overline{\rho_a^2},
\overline{\rho_m^2}$ are the average square radii of instantons
in the two components, $C$ is the normalization of the single 
instanton density (see 
(\ref{idens})) and $b=\frac{11}{3}N_c
-\frac{2}{3}N_f$ is the coefficient of the Gell-Mann-Low function.

   The model uses a simplified gluonic interaction corresponding
to an average repulsion $\langle S_{int}\rangle= \kappa \rho_1^2
\rho_2^2$ parameterized in terms of a single dimensionless constant
$\kappa$. The fermion determinant for the random component is 
approximated by $\langle \frac 1N {\rm Tr} TT^\dagger\rangle^{N_f/2}$ 
where the average is  over all positions and 
orientations. For the molecular component, on the other hand, the
overlap matrix element is first raised to the $N_f$ power and 
then averaged $\langle \left(T_{I\bar I}T_{\bar II}^\dagger
\right)^{N_f}\rangle$.  This
average is only performed over positions, whereas the relative orientation
is kept fixed. 

   Using the overlap matrix elements defined in the last section, one
finds a very specific temperature dependence of the quark induced 
interaction. The average determinant for the random component gradually 
decreases with temperature, whereas the determinant for the 
molecular component first increases (at $T\sim T_c$) and 
eventually, at larger $T$, starts to  decrease.
This behavior leads to the disappearance of random instantons 
at high temperatures and to a phase transition to a purely 
molecular phase in which chiral symmetry restored.

   As we are going to show in this section, the reason for this temperature 
dependence is given by a strong and rapid  polarization of the molecules
in the critical region. Polarization in this context means that
at finite $T$ the 
vector connecting the centers of the instanton and anti-instanton $R_\mu=
z_{\bar I \mu}-z_{I \mu}=(\vec r,\tau)$ is no longer distributed isotropically 
(as it is at $T=0$), but tends to point in the time direction.
Roughly speaking one may say that at $T\sim T_c$ the ``instanton liquid" 
actually becomes a ``liquid crystal" of the nematic type.

  In general, the anisotropy of a field theory  at finite 
temperature is of course a consequence of the different boundary conditions 
imposed in the time and spatial directions. An important consequence of
these boundary conditions is the fact that the propagators for bosons and 
fermions become qualitatively different at finite temperatures. Both are 
anisotropic, but fermions are subject to a qualitatively new phenomenon, 
the exponential screening of the propagator in the spatial direction. 
For a free massless fermion, the propagator is $S(r)\sim \exp(-\pi T r)$ 
where $\pi T$ is the lowest Matsubara frequency. Therefore, with increasing 
temperatures quarks develop a strong preference to propagate in the time 
direction. This is the main mechanism that produces an
anisotropy of the instanton ensemble at finite temperature \footnote{The 
gluonic $I\bar I$ interaction at finite T was  studied in \cite{SV_T}, 
with the result that it remains approximately isotropic until rather 
high temperatures.}.


   Before we come to quantitative results, let us give a qualitative 
explanation why the phenomenon takes place. Suppose the vector
 connecting the $I$ and $\bar I$ centers is $R_\mu=(\vec r,\tau)$. 
Using the overlap matrix elements specified above, the probability 
to have such a molecule is roughly proportional to
\be
\det(i\hat D)\sim |\sin(\pi T \tau)/\cosh(\pi T r)|^{2N_f}.
\ee
  This factor is maximal\footnote{Note also, that near the maximum
$\det(i\hat D)\sim \exp[-N_f\pi^2 T^2(\delta \tau^2+\delta r^2)]$, so at
large number of flavors one can use a saddle point method to evaluate
the contribution of this region.} for $r=0$ and $\tau=\beta/2=(1/2T)$, 
which is the most symmetric position of the $I\bar I$ pair on the 
torus. Note that it corresponds to a molecule polarized in 
the time direction. As a function of temperature one may expect the 
strongest effects to occur when the temperature is such that the period 
of the torus in the 
 direction is about twice the typical size of a molecule.
In the euclidean formalism, the phase transition becomes a simple geometric
effect: the transition occurs when the $I\bar I$ molecules optimally fit
into the 'temperature box'.

  In order to study this effect in more detail we have performed 
a simulation of the partition function (\ref{Z}) for a  single
$\bar I I$ pair at finite temperature $T$. 
The interaction is given by the fermion determinant discussed
in section 2 and the gluonic interaction specified in \cite{SV_T}.
The simulation is based on a simple Metropolis algorithm in
which a new random configuration of the $I\bar I$ pair is accepted 
with a probability $p\sim \exp(-S_{eff})$, where the effective action 
is the sum of the gluonic and fermionic parts, $S_{eff}=S_{int}+
\log(\det \hat D)$. Since there is only one pair present in our
simulation we keep the size of the instantons fixed while the 
positions and orientations are updated. 

  In order to characterize the orientation of the $I\bar I$ molecule 
we introduce the angle $\alpha$  between the vector $R_\mu$ 
(connecting the $I$ and $\bar I$ centers) and the (3-d) spacelike plane:
$\tan\alpha=R_0/|\vec R|=\tau/r$. The distribution of the
polarization is then determined by averaging the $\alpha$
over a large number of configurations. In fig. 1 we show the 
differential distribution of the angle $\alpha$ for different temperatures
and two different cases: (i) two light and one
heavy flavor (QCD); and (ii) four light flavors. The distributions are
normalized to the isotropic 4-d distribution, which means
they are divided by the relevant Jacobian $\cos^2\alpha$. In both cases 
 we observe that the distribution is practically flat
(=isotropic) at low temperatures $T<100$ MeV, but becomes strongly 
peaked at $\alpha=\pi/2$ for larger $T$. At very large temperatures,
this peak starts to disappear: this happens because at high $T$
 the time dimension is 
so strongly squeezed that no molecules can be oriented this way. As 
expected, the polarization is stronger in the case of four light
flavors. 

In fig. 2 we show fraction of molecules in the 
ensemble with $|\vec R| < {\rm min}(R_0, 1/T-R_0)$.
We observe that the degree  of polarization jumps to
a very large value $\simeq 70\%$ over a fairly small temperature
interval $T\simeq (120-150)$ MeV. 

  Finally, let us make a comment on the physical meaning of the 
polarization phenomenon in a less technical language. The $I\bar I$ 
molecules are virtual (or failed) tunneling events, in which the 
gauge fields penetrate into a new classical vacuum only for a short period 
of time, and then return back. For that reason, they do not contribute 
to the quark condensate and other related quantities. ``Polarization" of the 
$\bar I I$ molecules at $T\sim T_c$  means that 
at such temperatures the tunneling is concentrated in 
the vicinity of the same spatial point.

\section{Magnetic versus electric fields, and the equation of state}

Lattice data on the thermodynamics of the chiral phase transition
suggest that the pressure $p(T)$ remains relatively small around 
the transition region, while the energy density $\epsilon(T)$ 
changes by about an order of magnitude over a small\footnote{
For three and more massless quarks the transition is believed to be
of first order, but the phenomenon mentioned above is also observed 
for two massless quarks (when there is a second order phase transition)
and for light but  massive quarks, when there is no strict phase transition 
at all.} interval $\delta T \ll T_c$. A parametrization of the data 
obtained by Kogut et al. \cite{Kogut} together with a general discussion 
can be found in \cite{Koch_Brown}. These authors stress 
that at chiral restoration $T\sim T_c$ not only the (relatively small)
quark related energy and pressure are significantly changed, but that the 
(much larger) gluonic energy and pressure should also be modified.

Let us mention one more point emphasized in \cite{Koch_Brown}:
there seems to be a contradiction between the small transition 
temperatures $T_c\simeq 140$ MeV seen in unquenched lattice simulations 
and the large value of the phenomenological bag pressure $B\simeq
500\,{\rm MeV}/{\rm fm}^3$ derived from the trace anomaly 
\cite{Shuryak_conf,SVZ} (at $T$=0). With the usual bag  model 
equation of state, in which the pressure at $T>T_c$ is written 
as the sum of the Stefan Boltzmann contribution for an ideal 
quark-gluon gas minus the nonperturbative bag pressure, one is 
unable to get a positive pressure at such at low transition temperature! 
The natural alternative is to assume that the nonperturbative
phenomena responsible for the bag pressure at $T=0$ are partially present 
at $T>T_c$ as well, so that only part of this large $B$ is removed
across the transition region. 

   The question whether the rearrangement of the instanton ensemble 
into a molecular phase can explain this behavior was studied in
\cite{IS2}, where the schematic partition function defined by 
(\ref{molec},\ref{random}) was used to determine the instanton
related contribution to the equation of state. It was indeed found that 
instantons give a sizeable contribution to the energy density and 
pressure, that the two quantities behave quite differently and that
the results are roughly consistent with the lattice data. In particular, 
it was shown that, (i) $p_{inst}(T)$ jumps down, to about half of its 
value at zero temperature, while, (ii) $\epsilon_{inst}(T)$ jumps up, 
thus contributing to the ``latent heat" of the transition.

 The reason for this behavior was attributed to the specific 
temperature dependence of the fermion interaction used in that 
model. In the last section we have explained that this $T$-dependence
is caused by the  polarization of $I\bar I$ molecules.
Now we want to show, that this phenomenon leads to a simple
microscopic explanation of the energy density and pressure, 
related  to the (color) electric and magnetic fields associated 
with polarized molecules.

  Let us first recall some well known facts about instantons at $T=0$.
The instanton (anti-instanton) fields are selfdual (or anti-selfdual), 
$G_{\mu\nu}=\pm {1\over 2}\epsilon_{\mu\nu\alpha\beta}G_{\alpha\beta}$. 
In Minkowski space they correspond to classically forbidden paths
and the electric and magnetic fields are 
related by $E_k(x)=\pm i B_k(x)$. As a consequence, in the classical energy 
density $\epsilon_{cl} = (1/2)(E^2+B^2)$ the negative electric term is
exactly compensated by the positive magnetic one, so the configuration may
be created 'out of nothing', in agreement with the classical equations of 
motion.

  Furthermore, instantons contribute to the energy density of the QCD 
vacuum, but this happens not at the classical but at the quantum (one-loop)
level. This is quite natural, since instantons correspond to  tunneling 
phenomena, which are known to lower the ground state energy. In QCD, this 
fact can be expressed in terms of the  'stress tensor anomaly'  
\be 
\epsilon = -{b \over 128\pi^2} <(gG)^2>
\ee
which is non-zero for the 
instanton solution. Here, $b=(11/3)N_c-(2/3)N_f$ is the first coefficient of 
the beta function. For a dilute ensemble, one can directly relate
the energy density to the instanton density $\epsilon=-{b\over 4}n_{inst}$
\cite{Diakonov_Petrov}. As we already emphasized above, this energy is 
numerically rather large.

 At  finite instanton density (and especially in strongly correlated
molecules) the fields are not purely self-dual. However, at $T=0$
molecules are  unpolarized: therefore only Lorentz scalars can have 
non zero vacuum expectation values. Since the liquid is relatively dilute, 
these deviations from self-duality just lead to a relatively small
correction to the right hand side of the anomaly equation.

  At finite temperature, instantons contribute to the equation of
state in two ways. First, the instanton density will change, leading
to a change in the bag pressure. Second, the polarization of $I\bar I$
molecules in the time direction causes the expectation values of the 
electric and magnetic fields to be different. While the first effect 
gives equal but opposite contributions to the energy density and 
pressure, non-selfdual fields give a positive contribution to both
the energy density and the pressure. 

 In order to study this effect, we have calculated the color
fields for a single molecule with given polarization. Since $I\bar I$
configurations are not exact solutions of the euclidean
equations of motion the quantitative result depends on the specific ansatz 
for the gauge fields. Here we show the two most widely used forms of
the gauge potential, the sum ansatz \cite{Diakonov_Petrov} (full lines in
figs. 3 and 4) and the 
ratio ansatz \cite{Shuryak_88} (dashed lines in figs. 3 and 4).

In fig. 3 we show the modification of the action $S_{int}=S_{tot}-2S_0$ and the 
energy for a single molecule polarized in the time direction
at a temperature of 200 $MeV$. Results are given for the most attractive and
the most repulsive orientation. Indeed, one 
finds considerable attraction for the favored 
orientation $\cos\theta=1$. This interaction is weaker in the ratio ansatz, 
but the qualitative features are the same as in the sum ansatz. The new 
result is the lower panel, showing the polarization of the fields 
in the euclidean time direction.
Both in the sum and in the ratio ansatz cases, the fields
are  dominantly magnetic and give a positive contribution to the classical
euclidean energy density $\epsilon = 1/2\langle (B^2 - E^2)\rangle$. 
The effect is on the order of 15\% in the sum 
ansatz, whereas in the ratio ansatz it is roughly 10\%. 

In fig. 4 we show the same calculation for a molecule which is 
oriented along a spacelike direction. Again we observe attraction
for $\cos\theta=1$, but this attraction is considerably weaker 
as compared to a polarized molecule. Also, in the ratio ansatz 
the molecule is now essentially self dual, while in the sum ansatz 
it is actually dominantly electric.

 Let us now come back to the discussion of the energy density and
pressure
\be 
\epsilon &=& {1\over 2}\langle B^2-E^2\rangle - g^2{b \over 128\pi^2}
\langle E^2+B^2 \rangle,\\
 p &=&{1\over 6}\langle B^2 - E^2 \rangle + g^2{b \over 128\pi^2}
\langle E^2+B^2 \rangle .
\ee
The important point is that although $\langle E^2+B^2\rangle$ is 
much larger than $\langle B^2-E^2\rangle$, the former
quantity is suppressed by a smaller coefficient. Therefore, even
a moderate deviation from selfduality leads to a substantial contribution 
to the energy density and pressure! Taking $\langle B^2-E^2\rangle /
\langle B^2+E^2\rangle\simeq 10\%$ from the calculation of a single
molecule, one can get $\epsilon\simeq 225\,{\rm MeV}/{\rm fm}^3$ even
if the density of instantons does not change across the transition. 
Thus it appears that the observed jump in the energy
density can be explained by a change in the (instanton related) bag
pressure plus the contribution from instanton interactions and the
perturbative Stefan Boltzmann part.

\section{Quark interactions induced by $I\bar I$ molecules} 

  Instantons give a very successful phenomenology of the QCD vacuum, 
because they provide a mechanism for chiral symmetry breaking and
generate strong (and quite specific) interactions between 
light quarks. Although instantons do not lead to confinement 
the interactions are able to bind quarks into mesons and baryons 
and lead to a quantitative description
of the observed spin splittings. At the one-instanton level 
the interaction is described by the famous 't Hooft effective 
interaction \cite{tHooft}
\be
\label{thooft}
{\cal L} &\sim&  \prod_f (\bar\psi\phi_0)(\bar\phi_0\psi)
\ee
where $\psi$ are quark operators and $\phi_0$ is the zero mode
wave function. The product runs 
over the light quark flavors. This is still a complicated  nonlocal
interaction which depends on the color orientation of the instanton.
After taking the long wavelength limit and averaging over the 
color orientation it can be reduced to a local $2N_f$-point interaction
\cite{Shifman_et_al,Nowak_et_al}. In the case of two light flavors it
reads
\be
\label{leff}
  {\cal L}=  \int n(\rho)d\rho\, \frac{(2\pi\rho)^4}{8(N_c^2-1)} \left\{
     \frac{2N_c-1}{2N_c}\left[ 
     (\bar\psi\tau^{-}_a\psi)^2 + (\bar\psi\tau^{-}_a\gamma_5\psi)^2 
     \right]
     - \frac{1}{4N_c} (\bar\psi\tau^{-}_a\sigma_{\mu\nu}\psi)^2 \right\} ,
\ee 
where $n(\rho)$ denotes the density of instantons. Here, $\psi$ is an
isodoublet of quark fields and the  four vector $\tau^{-}_a$ has  
components $(\vec\tau,i)$ with $\vec\tau$ equal to the Pauli
matrices acting in isospace. The Dirac matrix $\sigma_{\mu\nu}$ is 
defined by $\sigma_{\mu\nu}=\frac{i}{2}[\gamma_\mu,\gamma_\nu]$.
The interaction (\ref{leff}) successfully  explains  many
properties of the ($T$=0) QCD correlation functions, most 
importantly the strong attraction seen in the pion channel.

  At finite temperature the collective coordinates of 
instantons are no longer random, but become
correlated. Therefore, it is no longer possible to average over the 
instanton orientations and use the effective Lagrangian (\ref{leff}).
However, due to the presence of molecule there should be strong  
non-perturbative effects even around  $T_c$ s.  For this reason 
we would like to investigate what kind of interactions are
induced by correlated $I\bar I$ pairs. We first study the quark 
propagator in the field of an instanton anti-instanton molecule 
and then derive the effective vertex that reproduces all four quark
correlations.

The quark propagator in the field of a single molecule is given by
\be
 S_{ij}^{ab}(x,y)&=& S_0(x,y) + \frac{1}{T_{I\bar I}} \left(
   \phi_{I,i}^a(x)\phi_{\bar I,j}^{b\,\dagger}(y) +
   \phi_{\bar I,i}^a(x)\phi_{I,j}^{b\,\dagger}(y) \right).
\label{prmol}
\ee
Here $\phi_I$ and $\phi_{\bar I}$ denote the zero mode wavefunctions
in the field of the instanton and the anti-instanton and $T_{I\bar I}$ 
is the corresponding overlap matrix element. For simplicity
we use the zero temperature profiles. We neglect all non-zeromode 
contributions except for the free propagator $S_0$. The propagator 
(\ref{prmol}) satisfies the following symmetry relation
\be
S^{\dagger}(x,y) = S(y,x).\label{prop1}
\ee
For $N_c =2$ the symmetry group is larger. In that case we 
can relate the propagator to its transpose by
\be
S(x,y) =  -  \sigma_2\gamma_5 C S^T(y,x)C\gamma_5 \sigma_2
\label{prop2}
\ee
which is also known as the Pauli-G\"ursey symmetry \cite{PG}. 
The Pauli matrix $\sigma_2$ acts in color space. A remnant of this 
symmetry is still present for more than two colors. If we remember 
that correlators of color singlet currents only depend on the
relative $I\bar I$ orientation, which in the case of molecules is given by
$R_\mu \tau_\mu^+$, it is clear that the symmetry (\ref{prop2}) holds in
the gauge that one of the pseudoparticles has the identity as orientation
matrix. In that case $\sigma_2$ contains a projector of $SU(N_c)$ on the
$SU(2)$ subgroup where the $I\bar I$ molecule has its support.

To investigate the consequences of the symmetries (\ref{prop1}, \ref{prop2})
let us consider the tensor decomposition of the propagator
\be
S^{ab}(x,y) = \sum_i a^{ab}_i(x,y) \Gamma_i,
\ee
where $\Gamma_i$ is the complete set of (hermitean) Dirac matrices.
Due to the chiral structure of the propagator only the vector
$a_{V\,\mu}$ and the axial vector $a_{A\,\mu}$ components 
are non-zero. For the scalar, pseudoscalar and tensor components
we have
\be
a_S = a_P = a_{T\,\mu\nu}  = 0.
\ee
As a consequence \cite{RILM} the scalar and the pseudo-scalar correlator
are degenerate. Also the axial and the vector correlator are degenerate, 
whereas the tensor correlator is not affected by the presence of molecules.

The symmetry (\ref{prop1}) leads to the relation
\be
a_i(x,y) = a^{\dagger}_i(y,x).
\label{prop4}
\ee
Under the restrictions quoted below (\ref{prop2}) also the following
two equalities hold
\be
a_{V\,\mu}(x,y) &=& -\sigma_2 a_{V\,\mu}^T(y,x) \sigma_2,\label{prop5}\\
a_{A\,\mu}(x,y) &=& \sigma_2 a_{A\,\mu}^T(y,x) \sigma_2,\label{prop6}
\ee
An immediate consequence of (\ref{prop5}) is that
\be
{\rm Tr} (S(x,x) \gamma_\mu) = 0. \label{prop7}
\ee
This term enters the disconnected contribution to the isoscalar correlation
function. The relation (\ref{prop7}) therefore implies that in
the molecular vacuum the rho and omega mesons are 
degenerate\footnote{This is quite remarkable since the near-degeneracy 
of the rho and omega mesons has no really good explanation. In fact,
finite-$T$ QCD sum rules suggest that the degeneracy is lifted at finite 
$T$ whereas our model predicts that it remains present.}. To exploit the 
relations (\ref{prop5},\ref{prop6}) further, we remind the reader the 
definition of a flavor non-singlet meson correlator
\be
\Pi^M_\Gamma(x,y)={\rm Tr } (S(x,y) \Gamma S(y,x) \Gamma ),
\ee
and the corresponding diquark correlator 
\be
\Pi^D_\Gamma(x,y) = {\rm Tr } (S(x,y) \Gamma C\sigma_2 S^{T}(x,y) 
        \sigma_2 C \Gamma ),
\ee
where we have chosen a particular color component of the diquark 
(every antisymmetric color matrix defines a diquark current). 
Again, $\sigma_2$ projects onto a $SU(2)$ subgroup of $SU(N_c)$.
From (\ref{prop2}) we conclude that meson correlators in the channel $\Gamma$
are degenerate with diquark correlators in the channel $\Gamma \gamma_5$
up to a color factor. This factor is given by $2/N_c(N_c-1)$ which counts
the total number of embedings of instantons in $SU(N_c)$. For $N_c=2$
the Pauli-G\"ursey symmetry becomes exact and meson and diquarks are 
degenerate. 

In \cite{RILM} it was shown that all meson correlators can be expressed
in the coefficients of the tensor decomposition of the propagator. In the
molecular vacuum, the coefficients obey simple relations which allow 
us to find relations between the correlators. We find
\be
 \overline{\sum_{kl}|a_{V\,\mu}^{kl}(x,y)|^2} = 3 
 \overline{\sum_{kl}|a_{A\,\mu}^{kl}(x,y)|^2},
\label{prop8}
\ee  
and
\be
 \overline{\sum_{kl}|a_{A\,\mu}^{kl}(x,y)|^2} = \frac 12 
 \overline{\sum_{kl}|a_{A\,\mu}^{kk}(x,x)a_{A\,\mu}^{ll}(y,y)|},
\label{prop9}
\ee
where the averaging is performed over the color and spatial 
orientation of the molecule and we have assumed that $x$ and
$y$ are far away from the locations of the pseudoparticles.
The relation (\ref{prop8}) determines unflavored correlation
functions \cite{RILM}, whereas (\ref{prop9}) gives the 
disconnected contribution for flavor singlet correlation functions.
Note that (\ref{prop7}) implies that $a^{kl}_{V\mu}(x,x)=0$. 
If the propagator is calculated in the spatial direction,
the relations (\ref{prop8},\ref{prop9}) remain valid even
the molecules are polarized in the temporal direction.

The effective vertex for quark-quark scattering in the molecular 
vacuum can be obtained using the standard reduction formula 
\be
<p_1p_2|{\cal L}_{mol}|p_3p_4>&=& - \int \prod_{i=1}^{4}d^4x_i\; 
    e^{ip_1x_1+ip_2x_2-ip_3x_3-ip_4x_4}
    \bar\psi^a_m(\hat p_1)_{mi}\bar\psi^b_n(\hat p_2)_{nj} \nonumber\\
    & & \hspace{0.2cm} <0|T(\bar q^a_i(x_1)\bar q^b_j(x_2)
    q^c_k(x_3)q^d_l(x_4))|0> (\hat p_3)_{kp} \psi^c_p
    (\hat p_4)_{lq} \psi^d_q ,
\ee
where $\psi^a_i$ is an isodoublet fermion field with color index $a$
and spinor index $i$. If we treat the correlated instanton ensemble
as a dilute gas of molecules, the vacuum expectation value can be 
calculated using the quark propagator (\ref{prmol}).
We only have to perform the  average over the global color orientation of 
a molecule whereas the relative orientation between the the instanton
and the anti-instanton is kept fixed. At low temperatures, molecules
are unpolarized and we average over the direction of the vector
connecting the centers of the instantons. After taking the long 
wavelength limit we obtain a local four fermion interaction
\be
\label{lmoleff}
 {\cal L}_{mol}&=& G \left\{ 
    \frac{1}{N_c(N_c-1)}\left[ 
        (\bar\psi\gamma_\mu\psi)^2+(\bar\psi\gamma_\mu\gamma_5\psi)^2 
        \right] 
        -\frac{N_c-2}{N_c(N_c^2-1)} \left[
        (\bar\psi\gamma_\mu\psi)^2-(\bar\psi\gamma_\mu\gamma_5
        \psi)^2 \right] \right. \nonumber \\
  & &  \hspace{0.4cm} +\frac{2N_c-1}{N_c(N_c^2-1)} \left[
        (\bar\psi\tau^a\psi)^2-(\bar\psi\tau^a\gamma_5
        \psi)^2 \right]\nonumber\\
  & &  \hspace{0.4cm} -\left.\frac{1}{N_c(N_c-1)}\left[ 
        (\bar\psi\tau^a\gamma_\mu\psi)^2+
        (\bar\psi\tau^a\gamma_\mu\gamma_5\psi)^2 
        \right]\right\}   
\ee
with the coupling constant
\be
 G = \int n(\rho_1,\rho_2)\,d\rho_1 d\rho_2\,
        \frac{1}{8T_{I\bar I}^2}(2\pi\rho_1)^2(2\pi\rho_2)^2\, .
\label{gmol}
\ee
Here $n(\rho_1,\rho_2)$ is the total tunneling probability for the 
$I\bar I$ pair. Note that the isospin structure of 
(\ref{lmol}) is different from (\ref{leff}), $\tau^a$ is a fourvector
with components $(\vec\tau,1)$. When calculating correlation functions
from this effective vertex both the direct and the exchange term have to
be taken into account. In order to extract the relevant coupling in 
each channel directly from the Lagrangian we add the exchange term 
and Fierz rearrange it into an effective direct interaction. The resulting
Fierz symmetric Lagrangian reads
\be
\label{lmol}
 {\cal L}_{mol\,sym}&=& G 
        \left\{ \frac{2}{N_c^2}\left[ 
        (\bar\psi\tau^a\psi)^2-(\bar\psi\tau^a\gamma_5\psi)^2 
        \right]\right. \nonumber \\ 
         & & - \;\,\frac{1}{2N_c^2}\left. \left[
        (\bar\psi\tau^a\gamma_\mu\psi)^2+(\bar\psi\tau^a\gamma_\mu\gamma_5
        \psi)^2 \right] + \frac{2}{N_c^2}
        (\bar\psi\gamma_\mu\gamma_5\psi)^2 \right\} + {\cal L}_8,
\ee
where ${\cal L}_8$ denotes the color octet part of the interaction
\be
\label{loct}
 {\cal L}_8&=& G \left\{
        \frac{N_c-2}{2N_c(N_c^2-1)} \left[
        (\bar\psi\tau^a\lambda^i\psi)^2-(\bar\psi\tau^a\lambda^i\gamma_5
        \psi)^2 \right] \right. \nonumber\\
   & &  \hspace{0.4cm}+ \frac{1}{4N_c(N_c-1)} \left[
        (\bar\psi\tau^a\lambda^i\gamma_\mu\psi)^2  +
        (\bar\psi\tau^a\lambda^i\gamma_\mu\gamma_5\psi)^2  
        \right] \nonumber \\
   & &  \hspace{0.3cm}- \left.\frac{1}{2N_c(N_c-1)} \left[
        (\bar\psi\lambda^i\gamma_\mu\psi)^2+
        (\bar\psi\lambda^i\gamma_\mu\gamma_5
        \psi)^2 \right]\right.\nonumber \\
   & &  \hspace{0.2cm}- \left.\frac{2N_c-1}{2N_c(N_c^2-1)} \left[
        (\bar\psi\lambda^i\gamma_\mu\psi)^2  -
        (\bar\psi\lambda^i\gamma_\mu\gamma_5\psi)^2  
        \right]\right\}.       
\ee

   At temperatures around the critical temperature for the chiral phase
transition, molecules become polarized in the time direction and we should 
not average over the relative position of the instanton and the anti-instanton.
In this case we get an effective lagrangian which is identical to 
(\ref{lmoleff},\ref{loct}) but with all vector interactions modified according
to $(\bar\psi\gamma_\mu\psi)^2\to 4(\bar\psi\gamma_0\psi)^2$.

 Like the effective interaction in the dilute instanton gas, (\ref{leff}),
the result for the molecular vacuum is $SU(2)\times SU(2)$ symmetric.
But in addition to being chirally symmetric, the effective lagrangian
in the molecular vacuum is also $U(1)_A$ symmetric. Local fermion
lagrangians of this type have been studied extensively in the context
of the Nambu and Jona-Lasinio (NJL) model \cite{njl}. However, in 
contrast to the NJL model the effective interaction in the instanton
model changes above the chiral phase transition. Furthermore, the 
NJL model has four unknown coupling constants, even when one 
restricts oneself to zero temperature and colour singlet terms only.
At finite temperature, the number of independent couplings is even 
larger since the space- and timelike vector currents can have 
different couplings constants. In our model, all coupling
constants are specified in terms of a single parameter $G$.

 Let us now compare the 't Hooft interaction (\ref{leff}) 
(dominant at $T=0$) and the 'molecule-induced' one we just derived, 
valid for $T>T_c$. The main effects of the 't Hooft interaction are (i) 
the strong attraction in the pion channel and (ii) the strong  repulsion 
in the $\eta'$ channel. The 'molecule-induced' effective lagrangian is also
attractive in the pion channel, but (since chiral and $U(1)_A$ symmetry 
are restored) the same attraction  also operates in the  scalar-isoscalar
($\sigma$),  scalar-isovector ($\delta$) and pseudoscalar-isoscalar ($\eta'$)
channels.

 Furthermore, this lagrangian also includes an attractive interaction in 
the vector and axial vector channels.
If molecules are unpolarized, the corresponding coupling 
constant is a factor of four smaller than the scalar coupling.
If they are completely polarized, the attraction is present only
in the operators containing $\gamma_0$, but  becomes equally strong 
as the attraction in the scalar-pseudoscalar channels.
Transversely polarized vectors are not affected in this case.

 There are further non-trivial qualitative features of the new effective 
Lagrangian. Note that it predicts no splitting between the isoscalar 
($\omega$) and isovector ($\rho$) vector channels. As we explained 
above, this is a consequence of the remnant of the Pauli-G\"ursey 
symmetry in the molecular vacuum. At the same time, there appears 
additional repulsion in the axialvector-isoscalar ($f_1$) channel 
compared to axialvector-isovector ($a_1$) case.

 In order to estimate the effective interaction in baryons we would like
to start with a simpler problem and study the interaction for diquarks.
The effective lagrangian (\ref{lmoleff}) can be rearranged into the form
\be
\label{ltrip}
 {\cal L}&=& G\left\{ \frac{2}{N_c^2(N_c-1)}\; 
                         \left[ (\bar\psi C\tau^A\beta^c\bar\psi^T)
                           (\psi^T C\tau^A\beta^c\psi)
                        -  (\bar\psi\gamma_5 C\tau^A\beta^c\bar\psi^T)
                           (\psi^T C\gamma_5 \tau^A\beta^c\psi^T)\right]
         \right.   \nonumber \\
    & & \hspace{0.2cm}-\frac{1}{2N_c^2(N_c-1)} 
                      \left[ (\bar\psi\gamma_\mu C\tau^S\beta^c\bar\psi^T)
                      (\psi^T C\gamma_\mu\tau^S\beta^c\psi)\right.
         \nonumber \\
    & &   \hspace{4cm} - \left. 
             (\bar\psi\gamma_\mu\gamma_5 C\tau^A\beta^c\bar\psi^T)
             (\psi^T C\gamma_\mu\gamma_5 \tau^A\beta^c\psi)\right]
    \Bigg\} + {\cal L}_6 
\ee
where ${\cal L}_6$ denotes the effective interaction among color sextet
diquarks. Only the anti-symmetric color matrices $\beta^c$ 
enter in (\ref{ltrip}). In order to facilitate the comparison with the mesonic
interaction \cite{Vogl}, they are normalized according to 
${\rm Tr}(\beta^c\beta^{c'})=N_c\delta^{cc'}$.
Using this normalization, the strength
of the interaction as compared to the free correlation function
can be inferred directly by  comparing the coupling constants.
For diquarks, the isospin wavefunction is determined by the dirac
structure \cite{RILM,Vogl}. Here, $\tau^S$ denote the
symmetric Pauli matrices and $\tau^A=\tau^2$ is the antisymmetric 
one. 

   From the effective lagrangian (\ref{ltrip}) we find an attractive
interaction for scalar $\psi^T C\gamma_5\psi$ and pseudoscalar
$\psi^T C\psi$ diquarks. The interaction for vector diquarks is 
also attractive with a coupling constant that is a factor 4
smaller than the one in the scalar channel. All coupling constants
are a factor $(N_c-1)$ smaller than the ones in the corresponding 
mesonic channel. Again, this is a consequence of the remnant of 
the Pauli-G\"ursey symmetry.

   The nucleon can be thought of as consisting of an equal mixture
of scalar and vector diquarks coupled to a third quark, whereas 
the delta resonance only contains vector diquarks. Using the 
effective interaction derived above, we would therefore expect
an attractive interaction for the nucleon. The delta-nucleon 
splitting, however, is expected to be significantly smaller than 
the pi-rho splitting since the coupling in the diquark channel is
a factor of 2 smaller than the coupling for mesons.

 The most direct consequence of the Lagrangian (\ref{lmoleff})
is the behavior of mesonic correlation functions. We will study
this question in the next section. The corresponding measurements
on the lattice provide the relevant test in order to check the
correctness of our model. Before we come to this, we would like 
to consider an even simpler set of physical quantities at nonzero 
$T$, the expectation values of four quark operators. These matrix
elements play an important role in QCD sum rule calculations and some
of them can be considered as  additional order parameters for chiral 
symmetry breaking. It is therefore interesting to check whether they 
vanish at $T=T_c$ in a universal manner. At low temperatures,
the temperature dependence of these expectation values can be determined
by calculating the contribution from single pion states.  At large
temperatures $T\simeq T_c$, however, little is known about the 
$T$-dependence. Usually it is assumed that the expectation values
can be factorized 
\be
 \langle\langle \bar q\Gamma q\bar q\Gamma'q\rangle\rangle &\simeq &
 \frac{1}{(4N_cN_f)^2} ({\rm Tr}(\Gamma){\rm Tr}(\Gamma') 
   -{\rm Tr}(\Gamma\Gamma') )\,\langle\langle\bar qq\rangle\rangle^2 ,
\ee
where the double brackets indicate thermal expectation values.
This would imply that all expectation values of four quark operators
are restored at $T_c$, and that they vanish at twice the rate of
the quark condensate. 

  Alternative $SU(N_f)$ order parameters are given by the expectation 
values 
\be 
O_1(T)&=& \langle\langle[\bar u_L \gamma_0 u_L-\bar d_L \gamma_0 d_L]
        [L\rightarrow R]\rangle\rangle,\\
O_2(T)&=& \langle\langle[\bar u_L \gamma_i u_L-\bar d_L \gamma_i d_L]
        [L\rightarrow R]\rangle\rangle,\\
O_3(T)&=& \langle\langle[\bar u_L \gamma_0 t^a u_L-\bar d_L \gamma_0 t^a d_L]
 [L\rightarrow R]\rangle\rangle,\\
O_4(T)&=& \langle\langle[\bar u_L \gamma_i t^a u_L-\bar d_L \gamma_i t^a d_L]
 [L\rightarrow R]\rangle\rangle ,
\ee
where $t^a=\lambda^a/2$ are the generators of $SU(N_c)$.
These operators enter the Weinberg sum rules at non-zero temperatures, 
related to the difference between vector and axial correlators 
\cite{Kapusta_Shuryak}. Instead of the operators listed above, one 
may also consider the combinations only containing left or right
handed quark fields. These operators are related to the sum of the 
vector and axial vector correlators, and need not be restored above
the chiral phase transition.

   A quantity sensitive to $U(1)_A$
violation is nothing else but the expectation value of the 't Hooft
vertex 
\be  
O^{'t \,Hooft}(T) &=& 
\langle\langle \det_f(\bar q_{f,R} q_{f,L}) \rangle\rangle, 
\ee
where the determinant is taken over the different quark flavors $f=u,d,
\ldots$ and $L,R$ stand for left and right components of the quark fields. 
Around or above $T_c$ there is no quark condensate, and 'unpaired'
instantons have very small density $O(m^{N_f})$ \cite{tHooft}. However,
in the measurement of quantities like this one, the denominators of
the quark propagators cancel the current quark masses, recovering
the famous 't Hooft effective interaction. In a sense, the operator 
considered can induce a tunneling event by itself. Unfortunately, 
this implies that a small fraction of the configurations will produce 
a large signal. This is certainly not an easy way for measurements to 
be performed.

   Finally, at the end of this section let us address the question of the
absolute strength of the 'molecule-induced' new effective interactions.
Instead of 7 unknown constants (which would appear in a general
NJL-type Lagrangian for $T>T_c$) we have only $G$, which is determined
by the number of $I\bar I$ molecules $n_{mol}(T)$.

Let us make a few general remarks. First, there should be a general 
upper bound,  $n_{mol}(T)< n_c$ (the density should be smaller than some 
critical number  for all $T>T_c$) following from the condition 
that the effective interaction should not be strong enough to lead to
a rearrangement of the vacuum  and cause spontaneous breakdown
of chiral symmetry. Second, in a fully polarized configuration the interaction
in the pion and longitudinal vector channel are identical. This might
lead to a situation in which all scalars as well as the longitudinal 
components of vector and axial mesons are massless. This speculative
scenario resembles the ``vector limit" studied in \cite{Georgi}.

Third, and the most natural case,
is that the density of molecules and the effective interaction is simply
too weak to produce any bound state. However, it can still be important and
lead to observable effects: one of them being modification of the 'screening 
masses' to be discussed in the next section.

\section{Correlation functions in the 'Cocktail Model'} 

   Having outlined the main qualitative features of the interactions
induced by $I \bar I$ molecules, and local averages induced by them,
 we now want to study more complicated
predictions of this model involving various correlation functions. 

   For this purpose we introduce a schematic model along the  
lines proposed in \cite{IS2}. In this model we assume that the total
number of instantons at temperatures around the chiral phase 
transition is not much smaller than at $T=0$. We then study the
phase transition by varying the fraction $f=2N_{mol}/N_{inst}$ of
instantons correlated in $\bar I I$ molecules. 

   We have carried out our simulations in a box $5.8^3\times 1.3
\,{\rm fm}^4$, corresponding to a total volume $V=256\,{\rm fm}^4$
and a temperature $T=150$ MeV. This is the temperature that
determines the boundary conditions on the fields and enters in the
calculation of the propagators. As we have explained in section 4 
we consider a scenario in which the phase transition takes place
over a fairly narrow interval of temperatures $\delta T \ll T_c$.
Since the instanton fields vary smoothly with temperature, we 
keep the size of the box fixed while we vary the fraction of
molecules.

   The calculation of the spectrum of the Dirac operator, the quark
condensates and hadronic correlation functions follows our earlier
work on the random instanton model \cite{RILM}. The essential idea
is to exactly diagonalize the Dirac operator in the space spanned
by the zero modes of the individual instantons and treat the effects
of the non zero modes on the single instanton level. In this work
we simply replace all the eigenfunctions and overlap matrix elements
by their finite temperature counterparts. This means in particular
that the quark propagator is antiperiodic in the time direction and
becomes screened for large spacelike separations.
   
   In fig. 5 we show the behavior of the quark condensate $\langle\langle
\bar qq\rangle\rangle$ (denoted by asterixes connected by the full line)
and the other vacuum expectation values introduced 
in section 5 as a function of the fraction $f$ of molecules. 
The dashed line shows the square of the condensate. 
The expectation values of the 't Hooft operator 
$\langle\langle O^{\rm 't\,Hooft}\rangle\rangle$ are represented by open 
squares. The behavior of the four quark condensates $O_1-O_4$
is almost indistinguishable (full squares) and we show only one of them.
All matrix
elements are normalized to their value at $f=0$. One clearly observes the 
restoration of chiral symmetry as the fraction of molecules approaches $f=1$. 
In our schematic model, not only chiral symmetry but also $U(1)_A$ symmetry 
is restored as $f\to 1$. This is apparent from the behavior of 
$\langle\langle O^{\rm 't\,Hooft}\rangle\rangle$, which is 
very large in the random model but vanishes in the molecular 
vacuum\footnote{As explained in section 5, this does not preclude the 
possibility that the $U(1)_A$ symmetry remains broken in an unquenched 
calculation.}.  The 
factorization assumption fails badly for this operator but
works fairly well for the vector operators $O_i$.


   More microscopic information is given by the spectrum of the Dirac
operator which we show in fig. 6. For the purely random vacuum, the 
spectrum is peaked at small eigenvalues $\lambda$. Using the Casher 
Banks formula $\langle\bar qq\rangle = -n({\lambda\!=\!0})/\pi$, 
this behavior shows that chiral symmetry is indeed broken. In the purely
molecular phase, the distribution of eigenvalues is peaked at finite
$\lambda$, the density of eigenvalues at zero vanishes and chiral 
symmetry is restored. The position of the maximum of the eigenvalue 
distribution reflects the typical inverse size of the molecules.
For arbitrary concentrations $f$ we observe that the spectrum looks
like a linear combination of a random and a molecular component.
This means that the Dirac operator really decomposes into a random 
part and independent $2\times 2$ blocks corresponding to molecules.

   Further details on the restoration of chiral symmetry are provided
by hadronic correlation functions. Here we will only discuss spacelike
correlators, since the corresponding screening masses have received 
a lot of attention in lattice calculations. We will present the 
corresponding temporal correlators in a forthcoming publication
\cite{SVS_97b}. The correlators in the pion, delta,
rho and $a_1$ channel are shown in fig. 7 and fig. 8. As the fraction of 
molecules approaches unity, one clearly observes that chiral symmetry is 
restored and the pion and delta as well as the rho and $a_1$ channels become
degenerate. However, this does not mean that the correlation functions
become purely perturbative. While the vector channels are indeed very 
close to the square of a massless thermal quark propagator, the scalar
meson correlation functions are significantly larger than what would 
be expected from the propagation of free quarks. 

    This is also apparent from the results for the spacelike screening 
masses which we show in fig. 10. If the correlation functions are dominated
by the propagation of free massless quarks, the spacelike screening masses
are expected to behave as $2\pi T$ for mesons and $3\pi T$ for baryons 
\cite{Eletski}. As the fraction of molecules approaches one, the screening
masses for the vector mesons $\rho$ and $a_1$ go to $2\pi T$ while the
pion remains significantly lighter. This is consistent with the effective
interaction discussed in the last section. Moreover, it is also in 
agreement with lattice results for the spacelike screening masses in
the relevant regime $T_c\sim 2T_c$ \cite{Gocksch}.

   We emphasize that the results shown in fig. 7~imply the restoration
of $U(1)_A$ as well as chiral symmetry \cite{PW} since the isovector scalar
($\delta$) meson is not the chiral partner of the pion. As was remarked
in \cite{Shuryak_ua1} this effect actually may have been observed on the 
lattice; the sigma meson correlation function determined in most lattice 
simulations is actually closer to the delta meson correlator. The reason 
is that the two loop graphs which distinguish the two correlation 
functions are  difficult to measure and  usually neglected in lattice 
simulations.

   In fig. 9 we show the behavior of the nucleon and delta correlation
functions. Among the various possible correlation functions we only 
show the chiral even correlators of the first Ioffe current for the 
nucleon $\eta_1=\epsilon^{abc}(u^aC\gamma_\mu u^b)\gamma_\mu\gamma_5
d^c$ and the standard delta current $j_\mu=\epsilon^{abc}(u^aC\gamma_\mu
u^b)u^c$ ($\Pi_2^N$ and $\Pi_2^\Delta$ in the notation of \cite{RILM}).
The chiral odd correlators vanish as $f\to 1$, in agreement with chiral
symmetry restoration. In this limit the chiral even correlators approach 
the cube of a thermal quark propagator. Similar to the pion channel, 
there is some residual interaction present in the nucleon channel. 
The screening mass of the delta approaches $3\pi T$ as $f\to 1$, whereas
the nucleon screening mass remains somewhat smaller. However, the effect
is much less pronounced as compared to the pion channel. 

As a consequence of
chiral symmetry restoration, the nucleon is expected to become degenerate 
with its chiral partner, which is usually identified with the lightest
odd parity $N^*$ resonance. Unfortunately, there is no current that
couples exclusively to the $N^*$ and not to the nucleon so that the 
corresponding screening mass is difficult to determine\footnote{In
lattice simulations one usually uses the fact that in staggered
fermion calculations different parity components are discretized on 
alternating lattice sites in order to identify the odd parity component
of the nucleon correlation function.}.  However, the fact that the 
chiral odd nucleon correlation functions vanish implies that any 
contribution from a positive parity nucleon resonance to these 
correlators is exactly cancelled by the contribution of the corresponding 
negative parity resonance. This provides at least an indirect check
for the statement that the nucleon and its chiral partner are 
degenerate.

\section{Conclusions} 
  
    In this paper we have studied various consequences of a scenario
in which the chiral phase transition in QCD is described as
a rearrangement of the instanton liquid, going from a random 
phase (at low $T$) to a correlated phase of polarized $I\bar I$ molecules 
at $T>T_c$. In this scenario, a significant number of  instantons is present
at temperatures $T=T_c\sim 2T_c$, causing a variety of
nonperturbative effects.

   As was shown in \cite{IS2} the phase transition does not require
a perturbative instanton suppression but can be triggered by the 
temperature dependence of the fermion determinant. Here we have
demonstrated that this particular behavior is due to the temperature
dependence of the fermion overlap matrix elements. At temperatures 
$T\simeq T_c$ these matrix elements strongly favor the appearance of
polarized instanton$-$anti-instanton molecules. This correlated 
instanton configuration leads to chiral symmetry restoration.

   One important effect caused by instantons is the significant
contribution to the energy density and pressure of the system
near the phase transition region.
 As demonstrated in section
4 the rapid polarization of the instanton molecules causes the 
energy density to jump up, while the pressure lags behind and remains
practically unchanged. This 
behavior is consistent with lattice simulations, in which the 
energy density rapidly reaches (or even overshoots) the perturbative
value whereas the pressure remains rather low until $T\simeq 2T_c$
\cite{Koch_Brown}. Thus, one can say that the energy of BNL AGS and CERN SPS
accelerators is now used to a significant degree in order to produce
gluomagnetic fields of  non-perturbative origin!

    The presence of $I\bar I$ molecules above $T_c$
also produces quite  specific interactions between light quarks. We
have derived an effective local $U(2)\times U(2)$ symmetric fermion
lagrangian which describes these effects. This Lagrangian is invariant
under a remnant of the Pauli-G\"ursey symmetry which results in the
degeneracy of the rho and omega channels.
We find that there is a
substantial residual attraction in the scalar-pseudoscalar meson channels even 
above $T_c$. This results is in agreement with lattice 
simulations, in which the presence of an attractive interaction in
the scalar channel has been established from an analysis of 
spacelike screening masses \cite{Gocksch} and direct measurements
of the scalar susceptibility \cite{Gupta}. 
The considerably weaker interaction in the vector channel 
follows from the behavior
of the quark number (vector) susceptibility \cite{Gottlieb,Kunihiro}.
In agreement with our effective Lagrangian, lattice calculations result
in a coupling constant that is four times larger in the scalar-pseudoscalar
channel than in the axial-vector channel \cite{Gupta,Boyd}.

    We have also established the presence of  molecule-induced
 interactions at $T\sim T_c$ from
direct measurements of the correlation functions in a schematic
'cocktail' model. In this model, the correlators are determined
in an instanton ensemble with a given fraction $f$ of correlated
$\bar I I$ pairs at temperatures around the critical
temperature $T\simeq 150$ MeV. As the instanton liquid becomes
completely correlated, we observe the restoration of chiral symmetry.
In this limit most of the spacelike screening masses approach their
perturbative values, $m=2\pi T$ for mesons and $m=3\pi T$ for baryons.
Notable exceptions are the scalar ($\pi,\sigma,\delta,\eta'$) mesons
whose screening masses are significantly smaller than $2\pi T$.

    Finally, we would like to comment that the 'mixed phase' 
(the region $T\sim T_c$) discussed above can be observed not only in
lattice simulations, but also in ongoing real experiments. In fact, 
it is exactly the kind of matter which is now produced and studied at 
the CERN SPS and Brookhaven AGS (to be complemented by future experiments 
at RHIC and LHC). The collective hydrodynamic expansion can naturally 
test the equation of state. Peaks in the dilepton spectra should provide
information on the masses of vector mesons decaying in matter, so 
hopefully one can check modifications of the masses of the rho, the
omega and the $a_1$ or the splitting of longitudinal and transverse 
components.

\section{Acknowledgements}
  The reported work was partially supported by the US DOE grant
DE-FG-88ER40388. Most of the calculations presented in this work
were performed at the NERSC at Lawrence Livermore. We have benefited from
many discussions about QCD thermodynamics with G.~E.~Brown and V.~Koch.
One of us (E.S.) would also like to acknowledge fruitful discussions with
E.-M.~Ilgenfritz and A.~Smilga.

\newpage\noindent
{\Large\bf figure captions}\\ \\ \\
\underline{figure 1} 
Distribution of the polarization angle of an 
instanton$-$anti-instanton molecule 
for different temperatures $T\,[{\rm MeV}]$.
The distributions are shown as a function of the orientation angle 
$\alpha$ introduced in the text and are normalized to a uniform
distribution $\cos^2\alpha$. The two figures show the two cases
of two light and one massive flavor (QCD) and four massless flavours
($N_f=4$).
\\ \\
\underline{figure 2}
Degree of polarization of an instanton anti-instanton molecule
as a function of the temperature. The two figures show the two cases
of two light and one massive flavor (QCD) and four massless flavors
($N_f=4$).
\\ \\
\underline{figure 3} 
Interaction and euclidean 'energy' $<E^2-B^2>$ in units of
the single instanton values for a molecule oriented in the timelike
direction at a temperature $T=200$ MeV. Results are given as 
a function of the molecule size $d$ in units of the instanton radius
$\rho=0.35$ fm. The dashed curves represents the results for the ratio 
ansatz whereas the full lines show the results for the sum ansatz. 
The lower line of each pair is for the most attractive orientation 
whereas the upper line is for the most repulsive orientation.
\\ \\
\underline{figure 4}
Interaction and euclidean 'energy' $<E^2-B^2>$ in units of
the single instanton values for a molecule oriented in the spacelike
direction at finite temperature $T=200$ MeV. Results are given as 
a function of the molecule size $d$ in units of the instanton radius
$\rho$. Curves as in figure 3. 
\\ \\
\underline{figure 5} 
Quark condensates in the 'cocktail' model for various concentrations
$f=2N_{mol}/N_{inst}$ with the total density $n_{inst}=0.75\,{\rm fm}^{-4}$
and the temperature $T=150$ MeV fixed. The various operators shown are 
defined in section 5 of the text. All expectation values are normalized
to their value at $f=0$. The solid and dashed lines shows the behavior of 
the quark condensate $\langle\bar qq\rangle$ and the square of the 
condensate $\langle\bar qq\rangle^2$, the open squares represent the 
expectation value of the 't Hooft vertex, and the expectation values
$<O_{1-4}>$ are denoted by the full squares.
\\ \\
\underline{figure 6}
Spectrum of the Dirac operator $i\hat D$ for various concentrations
$f=$0.00, 0.50, 0.75, 0.90, 1.00 of molecules. The density of instantons 
$n_{inst}=0.75\,{\rm fm}^{-4}$ and the temperature $T=150$ MeV are fixed. 
\\ \\
\underline{figure 7} 
Scalar meson correlation functions for various concentrations
$f=$0.00, 0.50, 0.75, 0.90, 1.00 of molecules. The density of instantons 
$n_{inst}=0.75\,{\rm fm}^{-4}$ and the temperature $T=150$ MeV are fixed.
The dashed curve shows the correlation function for massless quarks. 
\\ \\
\underline{figure 8}
Vector meson correlation functions for various concentrations
$f=$0.00, 0.50, 0.75, 0.90, 1.00 of molecules. The density of instantons 
$n_{inst}=0.75\,{\rm fm}^{-4}$ and the temperature $T=150$ MeV are fixed.
The dashed curve shows the correlation function for massless quarks. 
\\ \\
\underline{figure 9} 
Nucleon and delta correlation functions for various concentrations
$f=$0.00, 0.50, 0.75, 0.90, 1.00 of molecules. The density of instantons 
$n_{inst}=0.75\,{\rm fm}^{-4}$ and the temperature $T=150$ MeV are fixed. 
The dashed curve shows the correlation function for massless quarks.
\\ \\
\underline{figure 10}
Spacelike screening masses as a function of the concentration
$f$ of molecules. The density of instantons $n_{inst}=0.75\,{\rm fm}^{-4}$ 
and the temperature $T=150$ MeV are fixed. 

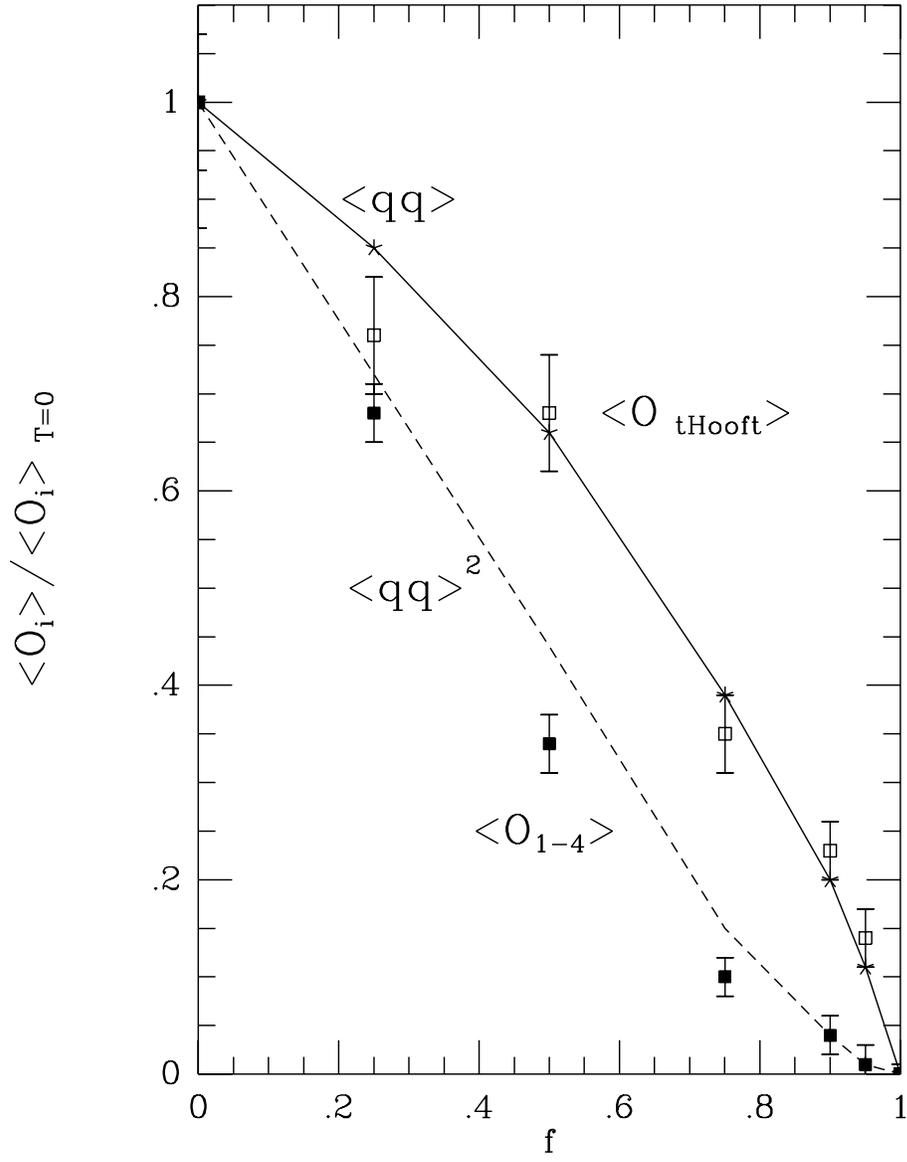
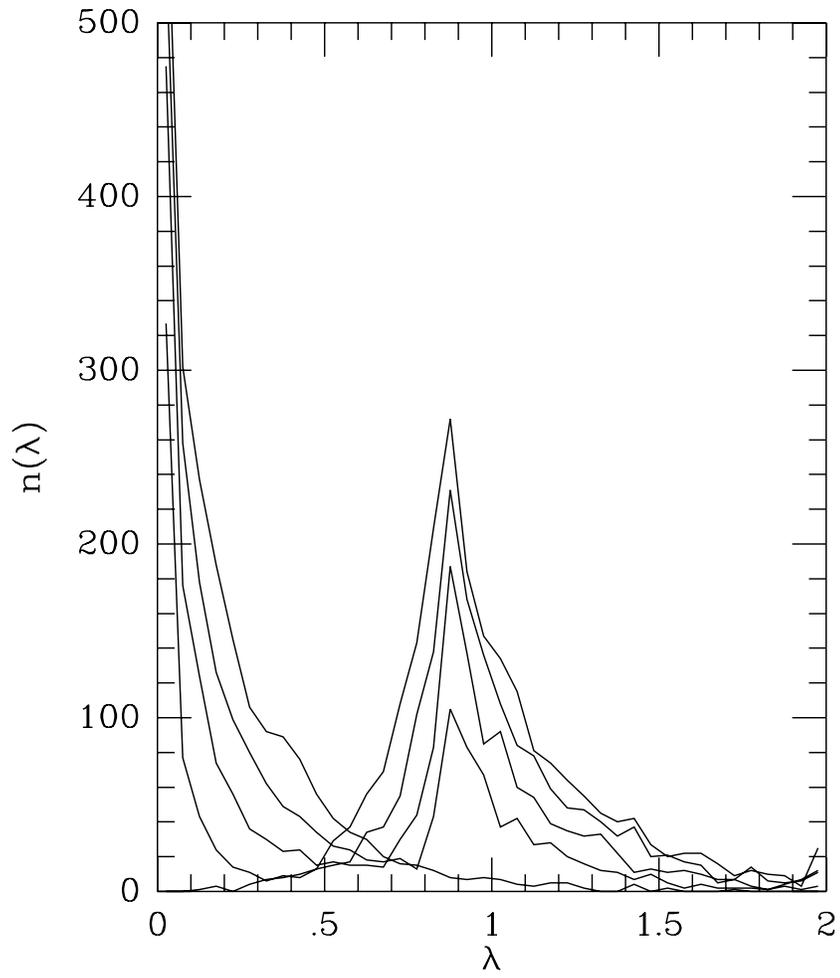
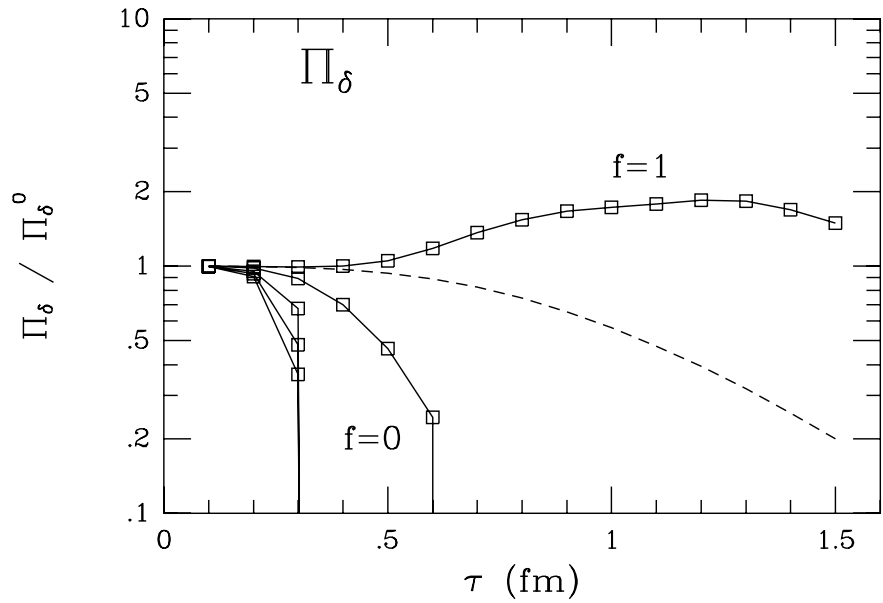
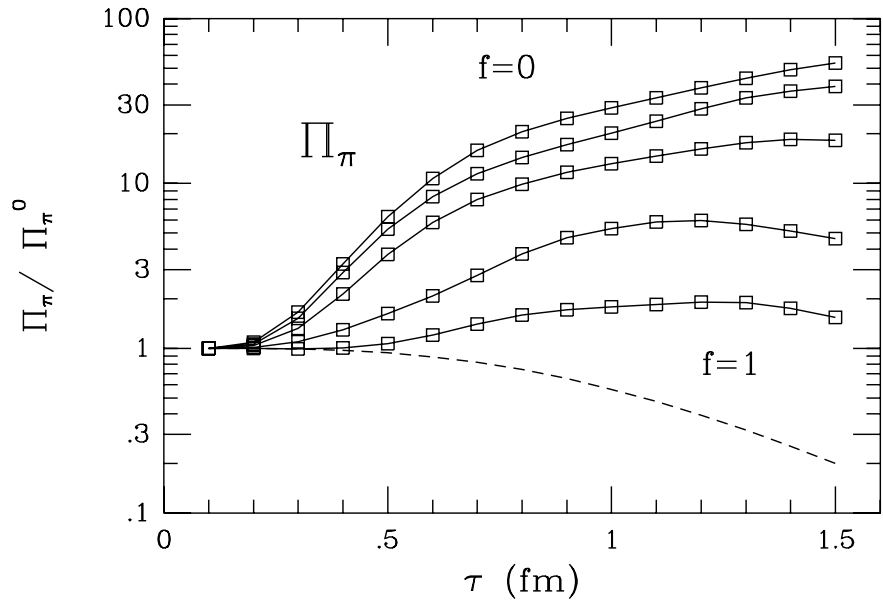
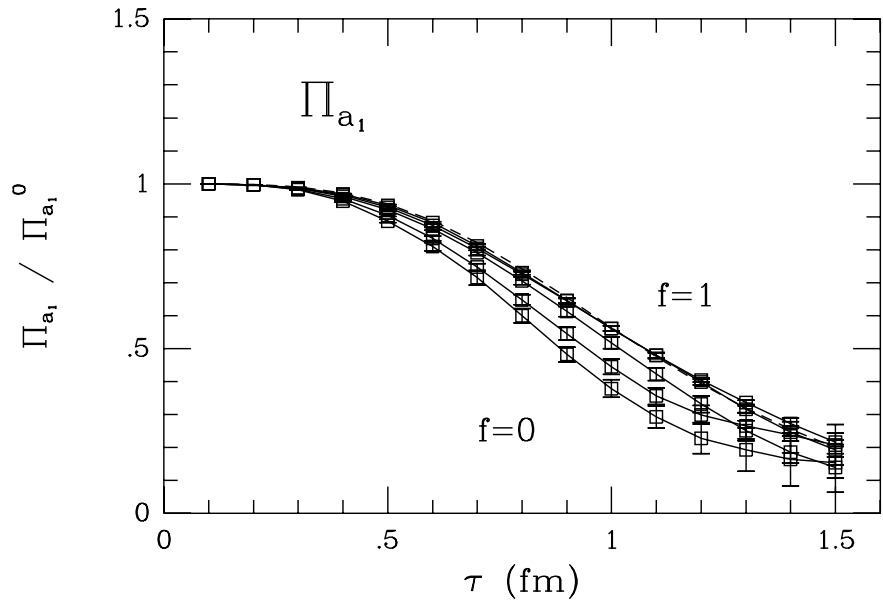
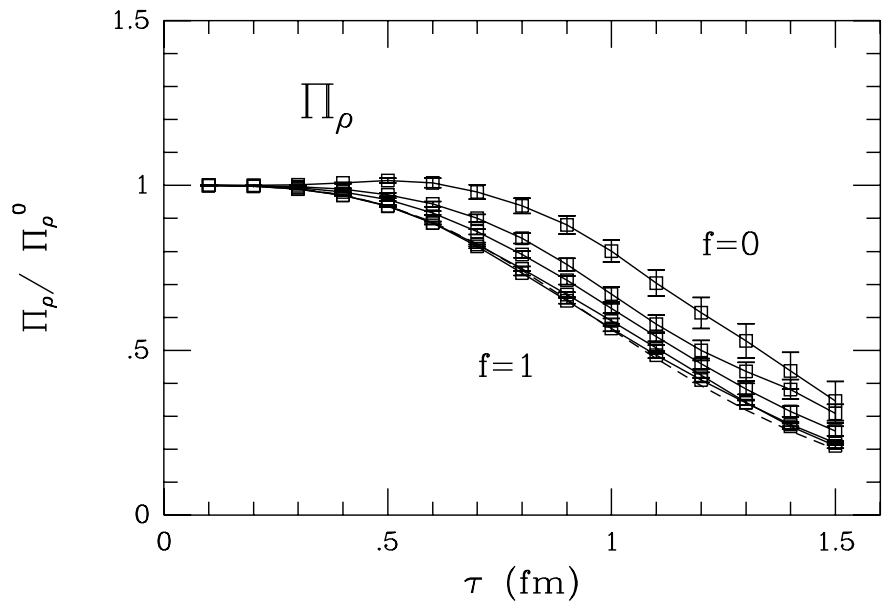
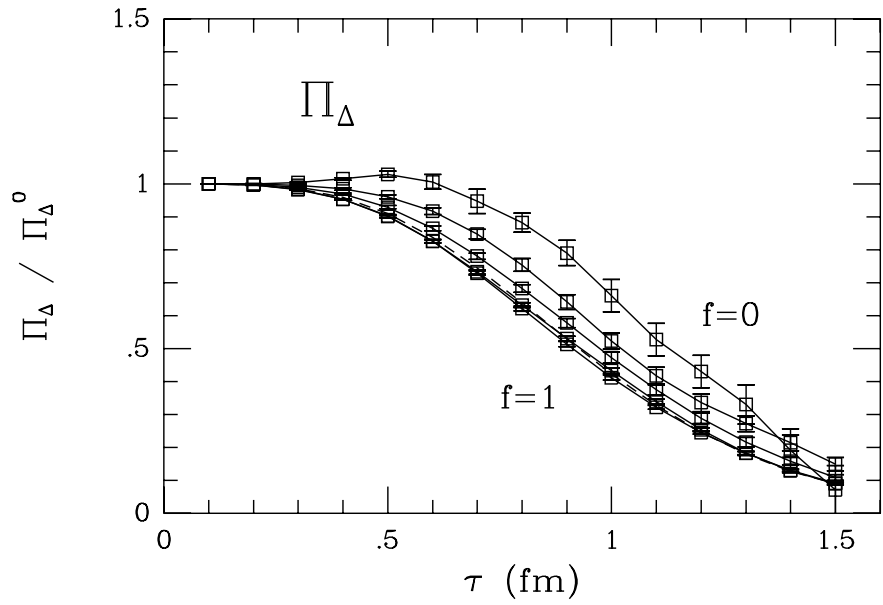
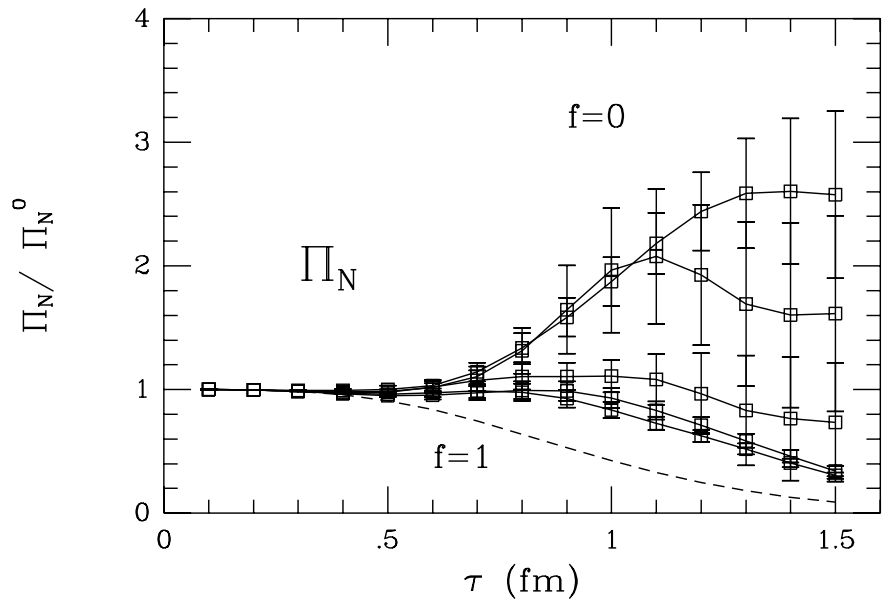
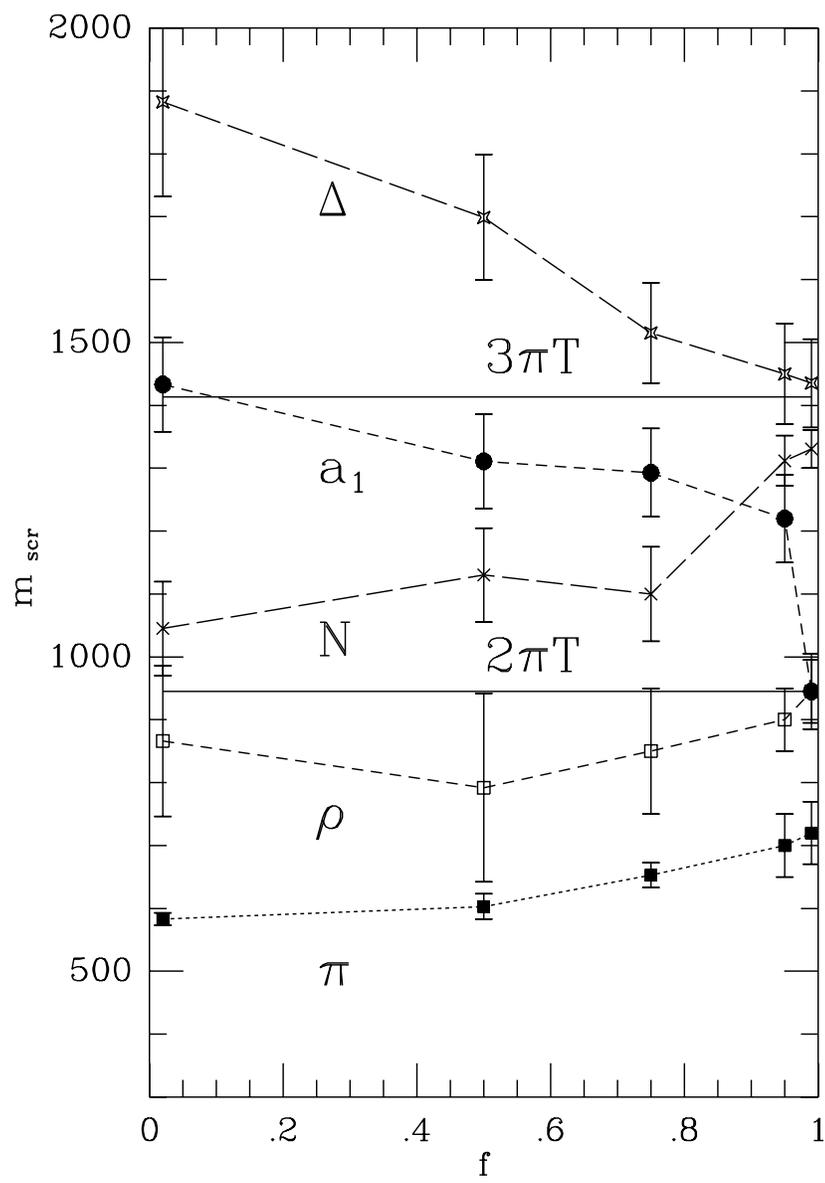
\end{document}